\documentclass[11pt]{article}

\usepackage[T1]{fontenc}
\usepackage{lmodern}
\usepackage[margin=1in]{geometry}
\usepackage{amsmath,amssymb,amsthm,mathtools}
\usepackage{microtype,booktabs,enumitem}
\usepackage{tikz}
\usetikzlibrary{calc}
\setlist{topsep=3pt,itemsep=2pt,parsep=0pt}
\usepackage[hidelinks]{hyperref}

\theoremstyle{plain}
\newtheorem{theorem}{Theorem}[section]
\newtheorem{lemma}[theorem]{Lemma}

\newtheorem{corollary}[theorem]{Corollary}
\theoremstyle{definition}
\newtheorem{definition}[theorem]{Definition}

\theoremstyle{remark}
\newtheorem{remark}[theorem]{Remark}

\newenvironment{procedure}[1]{%
  \par\medskip\noindent\hrule\vskip 2pt\noindent\textbf{#1}%
  \par\nobreak\vskip 2pt\begingroup\setlength{\parskip}{2pt}%
}{%
  \endgroup\vskip 2pt\hrule\par\medskip
}

\DeclareMathOperator{\aff}{aff}

\DeclareMathOperator{\lin}{lin}
\DeclareMathOperator{\vertx}{vert}
\DeclareMathOperator{\Int}{Int}
\DeclareMathOperator*{\E}{\mathbb{E}}
\DeclareMathOperator{\bits}{bits}
\newcommand{\Prb}{\mathbb{P}}
\newcommand{\R}{\mathbb{R}}
\newcommand{\one}{\mathbf{1}}

\newcommand{\pos}[1]{(#1)_{+}}
\newcommand{\Bn}{\mathcal{B}_n}
\newcommand{\Bnp}{\mathcal{B}_n^{+}}
\newcommand{\An}{\mathcal{A}_n}
\newcommand{\Ln}{\mathcal{L}_n}
\newcommand{\Sym}{\mathfrak{S}_n}
\newcommand{\mx}[1]{\lVert #1\rVert_{\max}}

\newcommand{\Round}{\textnormal{\scshape Round}}
\newcommand{\Mem}{\textnormal{\scshape Mem}}
\newcommand{\Lin}{\textnormal{\scshape Linear}}

\newcommand{\Fact}{\textnormal{\scshape Interior}}
\newcommand{\Univ}{\textnormal{\scshape Universal}}
\newcommand{\Amp}{\textnormal{\scshape Amplify}}
\newcommand{\dmin}{\delta_{*}}

\usepackage{xcolor}
\newcommand{\revcolor}[1]{#1}

\title{\bf Combinatorial Bernoulli Factories\\with Polynomially Many Coins}
\author{Renato Paes Leme\\ Google Research \and Rad Niazadeh\\ Chicago Booth}
\date{}

\begin{document}
\maketitle

\begin{abstract}
We provide an efficient implementation of combinatorial Bernoulli factories. Such a factory turns
coins of unknown biases $x$ into a random vertex of a polytope $P$ whose mean is \emph{exactly}
$x$. For every affine slice of the cube \revcolor{(the class of polytopes that admit a factory), we
give a single algorithm whose expected number of coin flips} and running time at every interior
$x$ are polynomial in the dimension and in $1/\dmin$, where $\dmin=\min_i\{x_i,1-x_i\}$ is never
supplied to \revcolor{the algorithm}. For perfect matchings this gives
$O(n^{8}\dmin^{-2}\log(n/\dmin))$ flips, while the factory of Niazadeh, Paes Leme\revcolor{,} and
Schneider spends \revcolor{at least} $n^{2n-1}$ \revcolor{flips in expectation} at the uniform doubly
stochastic matrix. \revcolor{For other polytopes, our method rounds exactly given a point deep inside
$P$ and the promise that $x$ is far from the boundary of $P$.}
\end{abstract}

\medskip
\section{Introduction}\label{sec:intro}

A \emph{Bernoulli factory} converts coins into coins\revcolor{: given} independent flips of a coin of
unknown bias $p$ and a known function $f$, \revcolor{it outputs} one bit that equals $1$ with
probability \emph{exactly} $f(p)$. The problem goes back to von Neumann's trick for extracting a
fair bit from a biased coin \cite{vonNeumann1951}\revcolor{. Keane and O'Brien
\cite{KeaneOBrien1994} characterized the functions $f$ that can be simulated, and a long line of
work has studied} how fast a factory can be
\cite{NacuPeres2005,MosselPeres2005,Huber2016,Huber2017,Latuszynski2011,Mendo2019,Morina2022}.

The central difficulty in the design of Bernoulli factories is to guarantee \emph{exact}
sampling. Approximate sampling is trivial \revcolor{(flip the coin $m$ times, form the empirical
mean $\hat p$, and output an $f(\hat p)$-coin), but unless $f$ is affine, its output is biased for
every $m$.} The applications also need exactness: it turns a black-box mechanism-design reduction
into one \revcolor{that preserves} incentive compatibility \emph{exactly} \cite{DHKN2021,CaiOVZ2021},
and it \revcolor{prevents errors from compounding} across iterations in exact simulation of diffusions
\cite{BeskosRoberts2005}, Monte Carlo \revcolor{methods} for intractable targets
\cite{FlegalHerbei2012,Latuszynski2011}, and quantum randomness processing
\cite{DaleJenningsRudolph2015}.

\medskip
Niazadeh, Paes Leme\revcolor{,} and Schneider \cite{NPS2023} introduced the combinatorial version. Fix a
polytope $P\subseteq[0,1]^{d}$ with vertex set $\vertx(P)$. One is given independent coins of
unknown biases $x_1,\dots,x_d$, promised that $x\in P$, and must output a random vertex $V$ with
\[
   \E[V]=x\qquad\text{exactly.}
\]
\revcolor{For example, when $P$ is} the Birkhoff polytope $\Bn$ of doubly stochastic
matrices\revcolor{, the problem is the following}: given coins \revcolor{for} the entries of an unknown
$X$, sample a perfect matching in which edge $(i,j)$ appears with
probability exactly $X_{ij}$. Their main theorem is a dichotomy: $P$ admits a Bernoulli factory
if and only if
\begin{equation}\label{eq:affineslice}
   P=\aff(P)\cap[0,1]^{d},
\end{equation}
that is, \revcolor{exactly when} $P$ is an affine slice of the cube. \revcolor{For the positive
direction, their factory} attaches to each vertex
$v$ a Bernstein polynomial $w_v$ with $\sum_v w_v(x)(x-v)=0$ on $P$ and runs a \emph{Bernoulli
race} \cite{DHKN2021}: propose a uniform vertex, run a finite one-bit factory for its polynomial,
accept on success and restart otherwise. A follow-up \revcolor{work} extends the explicit
constructions to
flow-based polytopes \cite{NPS2024}, and the multiparameter characterization of Paes Leme and
Schneider \cite{Multi2023} settles which functions of several unknown biases are simulable.

In this paper\revcolor{,} we address the main \revcolor{shortcoming} of \cite{NPS2023}\revcolor{, namely
its sample and computational inefficiency}. For $\Bn$ the race of \cite{NPS2023} takes the following form: pick a uniform
permutation $\pi$, flip the $n$ coins $X_{i\pi(i)}$, pick a uniform spanning tree of $K_n$
oriented toward vertex $1$, flip the $n-1$ coins $X_{i\pi(j)}$ for tree arcs $(i,j)$, and accept
only if all $2n-1$ flips are heads. The acceptance probability of a proposal is a product of
$2n-1$ entries of $X$\revcolor{. Even at the most benign input, the uniform matrix
$U=\tfrac1n\one\one^{\top}$,} every such product equals $n^{-(2n-1)}$, so the factory spends at
least $n^{2n-1}$ coin flips in expectation before it outputs anything. We give a factory whose
sample and computational \revcolor{complexities are} polynomial in $n$ and $1/\dmin$, where
$\dmin=\min_{ij}X_{ij}$ is the smallest entry of the unknown matrix, a quantity \revcolor{that is never
given to the factory} (Corollary~\ref{cor:matching}).

\revcolor{Our approach also extends} to polytopes beyond affine slices of the cube, under the promise
that the coin vector $x$ is far from the boundary of $P$\revcolor{, that is,} the $\ell_2$ ball of
radius $\rho$ around $x$, intersected with $\aff(P)$, lies in $P$ \revcolor{for} a supplied $\rho>0$. For example, given a point
deep inside $P$ and the promise that the coins are far from the boundary, we can design Bernoulli
factories for combinatorial polytopes such as the matroid base polytope and the perfect matching
polytope of a nonbipartite graph. Both come with polynomial-time membership and
vertex-decomposition procedures \cite{GLS1988}, and in general neither is an affine slice, so by
\eqref{eq:affineslice} a factory on the full domain need not exist. That factories exist on such
interior regions can be deduced from the multiparameter characterization of Paes Leme and
Schneider \cite{Multi2023}; we \revcolor{show} that they can be run with polynomially many coins and in
polynomial time.

\subsection{Our results}\label{sec:results}

Write $\mathbb B(x,r)$ for the $\ell_2$ ball of radius $r$ around $x$. For $\rho>0$ the
\emph{$\rho$-interior} of $P$ is
\[
   \Int_\rho(P)=\bigl\{x\in P:\ \mathbb B(x,\rho)\cap\aff(P)\subseteq P\bigr\},
\]
and $\Int(P)=\bigcup_{\rho>0}\Int_\rho(P)$ is the relative interior of $P$. Throughout, $d$ is
the dimension of the ambient cube and $T$ is the number of input-coin flips.

\paragraph{Interior exact rounding.} Our first result is a reduction. It converts two exact
deterministic \revcolor{oracles} for $P$ --- testing whether a rational point lies in $P$, and
decomposing a rational point of $P$ into vertices --- into exact sampling from coins, at a cost
\revcolor{that depends on} how deep inside $P$ the unknown point \revcolor{lies}. Given a rational point
$u\in\Int_r(P)$ together with $r$, and the promise that the unknown $x$ \revcolor{lies in}
$\Int_\rho(P)$ for a supplied $\rho$, there
is an algorithm that terminates almost surely, outputs
$V\in\vertx(P)$ with $\E V=x$ exactly, and uses
\[
   \E[T]=O\!\left(\frac{d^{7/2}}{r\,\rho^{2}}\log\frac{d}{r\rho}\right)
\]
input-coin flips \revcolor{and} $O(d^{3/2}/r)$ membership tests \revcolor{in expectation, plus a single
vertex decomposition}, with expected running time polynomial in $d$, $1/r$, $1/\rho$\revcolor{,} and
the encoding lengths (Theorem~\ref{thm:interior}). \revcolor{This result does not rely on}
\eqref{eq:affineslice}: the reduction applies to every rational polytope, and the promise
$x\in\Int_\rho(P)$ replaces the affine-slice condition.

\revcolor{The idea behind this result is to batch and correct. We project the empirical mean of a
batch of coin flips onto $\aff(P)$; the projected point has expectation exactly $x$ for every
batch size, but it may fall outside $P$. On this rare event, instead of discarding the sample, we
keep the signed residual and add its expectation back through a lottery over known feasible
points; the coin that drives this lottery is generated by a scalar Bernoulli factory.
Concentration is used only to certify that the correction is rarely needed, which keeps the
scalar factory within its promise; it never enters the computation of the output mean.}

\paragraph{Universal factories on affine slices.} When $P=\aff(P)\cap[0,1]^{d}$\revcolor{,} the promise
can be dropped. \revcolor{For such a $P$, a valid interior radius at $x$ is given by the coordinates of
$x$: $x\in\Int_{\dmin}(P)$ for $\dmin=\min_i\{x_i,1-x_i\}$ (Lemma~\ref{lem:cuberadius}). So this
radius is a statistic of the coins rather than outside information, and a pilot stage can
estimate it. However, an exact algorithm cannot trust any such estimate
(Section~\ref{sec:nomargin}). Instead, we make the reserve (the probability with which the
algorithm runs the correction) \emph{random}, and amplify the correction by the reciprocal of the
mean reserve, which is unknown; the pilot's mistakes are then paid for by the same reserve that
pays for the residual. We obtain a single algorithm that receives $P$ and nothing about $x$, and
that}
for every $x\in\Int(P)$ terminates almost surely, outputs $V\in\vertx(P)$ with $\E V=x$ exactly,
and uses
\[
   \E[T]=O\!\left(\frac{d^{7/2}}{\dmin^{3}}\log\frac{d}{\dmin}\right)
\]
input-coin flips, in expected time polynomial in $d$, $1/\dmin$\revcolor{,} and the encoding length of
$P$ (Theorem~\ref{thm:universal}). By \eqref{eq:affineslice}\revcolor{, affine slices form} the entire
class of polytopes that admit a factory. \revcolor{Our construction follows Morina's view \cite{Morina2021} of a Bernoulli factory as an unbiased estimator obtained by
debiasing a biased one, and it divides by the unknown mean reserve using the two-coin algorithm of
Gon\c{c}alves, \L{}atuszy\'nski, and Roberts \cite{GLR2017}.} For perfect matchings\revcolor{, this}
gives
$O(n^{8}\dmin^{-2}\log(n/\dmin))$ flips with $\dmin=\min_{ij}X_{ij}$
(Corollary~\ref{cor:matching}), an exponential improvement over the $n^{2n-1}$ flips that the
race of \cite{NPS2023} spends at $U$.

\subsection{Related work}

\paragraph{Scalar factories.} Von Neumann \cite{vonNeumann1951} started the subject \revcolor{by
showing how to extract} a fair bit from a biased coin. Keane and O'Brien \cite{KeaneOBrien1994}
then characterized which $f$ admit a factory at all, leaving open how fast \revcolor{a factory} could
be\revcolor{. Nacu and Peres \cite{NacuPeres2005} answered this question by} characterizing fast
simulability and giving the first fast constructions, and
Mossel and Peres \cite{MosselPeres2005} studied what finite automata can simulate.
\L{}atuszy\'nski et al.\ \cite{Latuszynski2011} gave a general reverse-time martingale scheme, and
attention then turned to the workhorse primitive $p\mapsto cp$\revcolor{. Huber}
\cite{Huber2016} gave a
nearly optimal linear factory using $\E[T]\le9.5\,c/\varepsilon$ flips under the promise
$cp\le1-\varepsilon$, with a matching lower bound up to constants, and later treated small means
\cite{Huber2017}; Mendo \cite{Mendo2019} did the same for power series. Morina et al.\
\cite{Morina2022} went further, simulating general rational functions of a categorical
distribution by perfect sampling of Markov chains, and Morina's thesis \cite{Morina2021} recast
the factory problem as the debiasing of an estimator, a view developed further by Koskela,
Karvonen, \L{}atuszy\'nski\revcolor{,} and Span\`o \cite{Koskela2025}. We use a linear factory as a
black box\revcolor{; for completeness, Appendix~\ref{app:scalar} gives a self-contained
implementation.}

\paragraph{Combinatorial factories.} The polytope formulation, the characterization
\eqref{eq:affineslice}, and the explicit bipartite matching factory are due to \cite{NPS2023};
\cite{NPS2024} extends the explicit constructions to flow-based polytopes and \cite{Multi2023}
gives the multiparameter characterization, including inputs on the cube boundary. For the
$k$-uniform matroid the construction of \cite{NPS2023} recovers Sampford sampling
\cite{Sampford1967}, and they show that for $1<k<d-1$ no \emph{strong} factory for $k$-subsets has
tails that decay exponentially, uniformly over the polytope. None of this work bounds the number
of coin flips.

\paragraph{Optimization and mechanism design.} We use the constructive Carath\'eodory
decomposition of Gr\"otschel, Lov\'asz\revcolor{,} and Schrijver \cite[Thm.~6.5.11]{GLS1988} as a
black box \revcolor{to decompose a rational point of an affine slice of the cube into vertices. This
step is standard; our contribution is to show that, together with coin access, it suffices for
\emph{exact} sampling.} Bernoulli factories entered theoretical computer science through Dughmi,
Hartline, Kleinberg\revcolor{,} and Niazadeh \cite{DHKN2021} (see also \cite{CaiOVZ2021}), where
\revcolor{each coin flip is simulated} by running an allocation algorithm, so polynomial coin
complexity makes their reduction efficient.

\paragraph{Sampling with prescribed marginals.} A large literature \revcolor{studies how to sample}
combinatorial objects with given marginals, via maximum entropy, counting\revcolor{,} or Markov
chains\revcolor{. This line of work assumes that the marginals are \emph{known} and settles for
approximate marginals; in our setting, the marginals are unknown and must be matched exactly. We
make no claim about \emph{which} distribution over vertices with mean $x$ is produced; in
particular, we do not claim any maximum-entropy, product-weight, or negative-dependence
property.}

\subsection{Organization}

Section~\ref{sec:prelim} fixes notation and the oracle model. Section~\ref{sec:idea} describes
the algorithm behind Theorem~\ref{thm:interior}, the geometry that makes it work, and its proof.
Section~\ref{sec:nomargin} removes the \revcolor{need for a supplied $\rho$ on} affine slices of the
cube, proves
Theorem~\ref{thm:universal}, and derives the bound for perfect matchings
(Corollary~\ref{cor:matching}).
Appendix~\ref{app:scalar} \revcolor{gives a self-contained implementation of the scalar linear factory,
the only probabilistic primitive that we use as a black box.}
\section{Preliminaries}\label{sec:prelim}

\paragraph{Polytopes.} A rational polytope is the convex hull of finitely many points of
$\mathbb Q^{d}$; equivalently, it is the bounded solution set of a finite system of linear
inequalities with rational coefficients. Throughout, $P\subseteq[0,1]^{d}$ is a nonempty rational
polytope with vertex set $\vertx(P)$. Its space of directions is $L=\lin(P-P)$, the linear span of
the differences of points of $P$, and its affine hull is $\aff(P)=u+L$ for any $u\in P$. We
write $\Pi$ for the orthogonal projection of $\R^{d}$ onto $\aff(P)$. It is an affine map that
fixes $\aff(P)$ pointwise, and its linear part $\Pi_L(w)=\Pi(u+w)-u$ is the orthogonal projection
onto $L$. We assume a rational description of $\aff(P)$ is available, so that $\Pi$ is computable
by rational linear algebra. For $t\in\R$ we write $\pos t=\max(t,0)$. Vectors are compared
coordinatewise.

\paragraph{Interiors.} Let $\mathbb B(x,r):=\{z:\|z-x\|_2\le r\}$ be the $\ell_2$ ball of radius
$r$ around $x$. For $\rho>0$,
\[
   \Int_\rho(P)=\bigl\{x\in P:\ \mathbb B(x,\rho)\cap\aff(P)\subseteq P\bigr\}
\]
is the $\rho$-interior of $P$, and $\Int(P)=\bigcup_{\rho>0}\Int_\rho(P)$ is its relative
interior. Throughout, $\dmin=\min_i\{x_i,1-x_i\}$ is the $\ell_\infty$ distance from the unknown $x$ to
the boundary of the cube. When $P=\aff(P)\cap[0,1]^{d}$\revcolor{,} we assume that no coordinate is
constant on $P$ \revcolor{(a constant coordinate can be detected by linear programming and deleted),}
so that
$\Int(P)=P\cap(0,1)^{d}$, and then $x\in\Int_{\dmin}(P)$ (Lemma~\ref{lem:cuberadius}).

\paragraph{Matchings.} \revcolor{We write} $\Sym$ \revcolor{for} the symmetric group on $[n]$ and $M_\pi$
\revcolor{for} the permutation matrix
of $\pi\in\Sym$. The Birkhoff polytope and its relative interior are
\[
  \Bn=\{X\in\R_{\ge0}^{n\times n}:X\one=\one,\ X^{\top}\one=\one\},
  \qquad
  \Bnp=\{X\in\Bn:X_{ij}>0\ \forall i,j\},
\]
and $\vertx(\Bn)$ consists of the $n!$ permutation matrices. We write $U=\one\one^{\top}/n$,
$H=I-U$, $\mx A=\max_{ij}|A_{ij}|$, and
\[
  \An=\{A:A\one=\one,\ A^{\top}\one=\one\},\qquad
  \Ln=\{A:A\one=0,\ A^{\top}\one=0\},
\]
so that $\An=\aff(\Bn)$ and $\Ln$ is its direction space.

\paragraph{Input model.} The input is a family of independently reusable coins with biases
$x_1,\dots,x_d$, where $x\in P$ is unknown; every query uses a fresh independent sample. The
algorithm may also flip auxiliary coins of known rational bias\revcolor{; such coins can be
implemented} from independent fair bits at expected bit cost polynomial in \revcolor{their} encoding
length, \revcolor{and we do not count their flips as input-coin flips}. We report (a) the expected number of input-coin flips, (b) the expected number
of oracle calls, and (c) the expected number of bit operations.

Following \cite{NPS2023}, a Bernoulli factory is a rooted binary decision tree whose internal
nodes are labeled by an input coin or by a known constant $c\in(0,1)$ and whose leaves are
labeled by outputs. A leaf of such a tree is reached with probability
\begin{equation}\label{eq:bernstein}
   c\prod_{i=1}^{d}x_i^{a_i}(1-x_i)^{b_i}
\end{equation}
for a constant $c\ge0$ and nonnegative integers $a_i,b_i$\revcolor{; that is, this probability is a Bernstein monomial.
This representation is the only feature of the model that we use, and we use it only in}
Section~\ref{sec:nomargin}.

\paragraph{Exact oracles.} We assume two exact deterministic procedures for $P$:
\begin{itemize}[leftmargin=1.4em,itemsep=1pt,topsep=3pt]
\item $\Mem(z)$: given a rational $z$, decide whether $z\in P$;
\item $\Round(q)$: given a rational $q\in P$, compute a rational convex decomposition
  $q=\sum_{j=1}^{\ell}\lambda_j v_j$ with $v_j\in\vertx(P)$, and return $v_j$ with probability
  $\lambda_j$.
\end{itemize}
\revcolor{We assume that both run in time polynomial in the encoding length of their input, in the
oracle model in which $P$ is given.} Note that $\Round$ is \emph{mean preserving}: $\E[\Round(q)]=q$. For
$0/1$ polytopes both are supplied by strong linear optimization
\cite[Thm.~6.5.11]{GLS1988}.

\paragraph{The linear factory.} We use one probabilistic primitive.

\begin{definition}[Linear factory]\label{def:linear}
$\Lin(c,\text{$p$-coin})$ takes a rational $c\ge2$ and independent flips of a $p$-coin with the
promise $cp\le\tfrac14$, and outputs a bit of mean exactly $cp$. We write $K_0$ for a constant with
$\E[\#\text{$p$-coin flips}]\le K_0c$.
\end{definition}

Huber's factory \cite{Huber2016} uses at most $9.5\,c/\varepsilon$ flips \revcolor{in expectation}
under the promise $cp\le1-\varepsilon$\revcolor{. Our promise $cp\le\tfrac14$ allows
$\varepsilon=\tfrac34$, which gives} $K_0\le38/3$.
Appendix~\ref{app:scalar} gives a self-contained non-optimized alternative with $K_0\le140$. Both
terminate almost surely, remain valid when $p=0$, and have expected bit cost polynomial in $c$
and $\bits(c)$. All our statements are in terms of $K_0$.

\section{Interior exact rounding}\label{sec:idea}

This section describes the algorithm behind our first result, the geometry that makes it work,
and its analysis. Throughout, $P\subseteq[0,1]^{d}$ is a rational polytope with the oracles $\Mem$ and
$\Round$ of Section~\ref{sec:prelim}; we are given a rational point $u\in\Int_r(P)$ together with
$r$, and we are promised that the unknown coin vector $x$ lies in $\Int_\rho(P)$ for a supplied
$\rho$, where $r,\rho\in(0,1]$ are rational. \revcolor{We use $s=\lceil\sqrt d\rceil$ as a rational stand-in for $\sqrt d$ wherever one is
needed, so that the biases of the coins that the algorithm flips are rational.}

\subsection{Batch and project}\label{sec:batchproject}

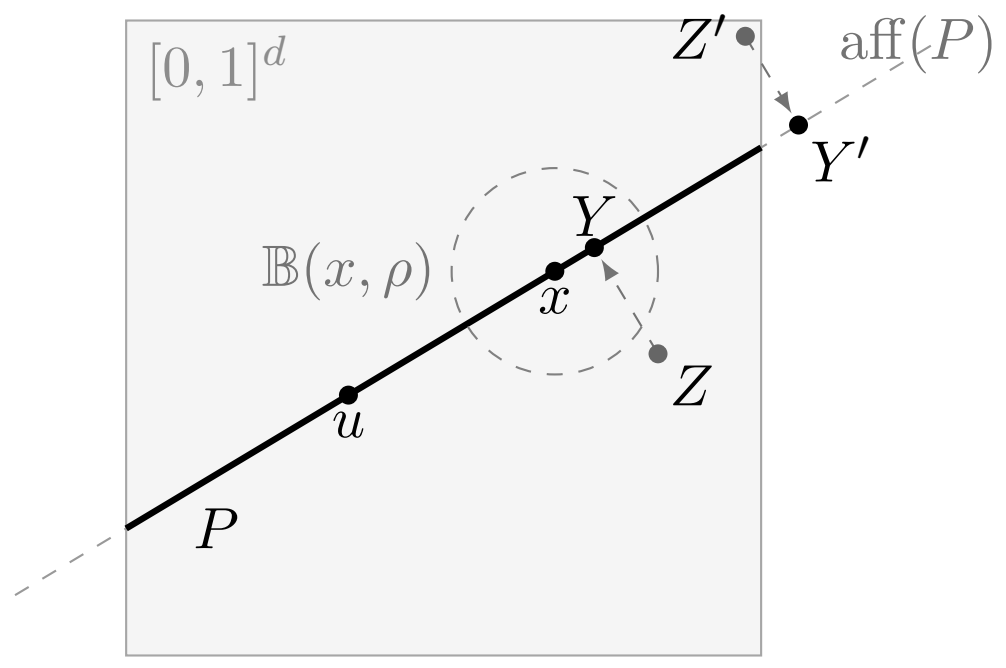
\begin{figure}[t]
\centering
\begin{tikzpicture}[scale=1.05,>=latex]
  \draw[fill=black!4,draw=black!35] (0,0) rectangle (4,4);
  \node[black!45,anchor=north west,inner sep=2pt] at (0.05,3.98) {$[0,1]^{d}$};
  \draw[black!40,dashed] (-0.7,0.38) -- (5.1,3.86);
  \node[black!55,anchor=west,inner sep=2pt] at (4.42,3.85) {$\aff(P)$};
  \draw[very thick] (0,0.8) -- (4,3.2);
  \node[anchor=north west,inner sep=2pt] at (0.35,1.0) {$P$};
  \coordinate (u) at (1.4,1.64);
  \coordinate (x) at (2.7,2.42);
  \coordinate (Z) at (3.35,1.9);
  \coordinate (Y) at (2.949,2.569);
  \coordinate (Zp) at (3.9,3.9);
  \coordinate (Yp) at (4.235,3.341);
  \draw[dashed,black!50] (x) circle (0.65);
  \node[black!55,anchor=east,inner sep=2pt] at ($(x)+(180:0.68)$) {$\mathbb B(x,\rho)$};
  \draw[->,black!55,dashed,shorten >=2.5pt] (Z) -- (Y);
  \draw[->,black!55,dashed,shorten >=2.5pt] (Zp) -- (Yp);
  \fill (u) circle (1.7pt); \node[anchor=north,inner sep=3pt] at (u) {$u$};
  \fill (x) circle (1.7pt); \node[anchor=north,inner sep=3pt] at (x) {$x$};
  \fill[black!60] (Z) circle (1.7pt); \node[anchor=north west,inner sep=2pt] at (Z) {$Z$};
  \fill (Y) circle (1.7pt); \node[anchor=south,inner sep=2pt] at (Y) {$Y$};
  \fill[black!60] (Zp) circle (1.7pt); \node[anchor=east,inner sep=3pt] at (Zp) {$Z'$};
  \fill (Yp) circle (1.7pt); \node[anchor=north west,inner sep=2pt] at (Yp) {$Y'$};
\end{tikzpicture}
\caption{The batch, drawn for $d=2$ with $\aff(P)$ a line. The empirical mean $Z$ of $m$ rounds of
flips need not lie on $\aff(P)$; its projection $Y=\Pi(Z)$ does, and $\E Y=x$ exactly. The ball
$\mathbb B(x,\rho)\cap\aff(P)$ lies in $P$ and $Z$ concentrates around $x$, so $Y$ is feasible
except on a rare event, as for $Z'$; the algorithm then rounds $u$ instead, and keeps the residual
$Y'-u$ rather than discarding the sample.}
\label{fig:batch}
\end{figure}

The natural first attempt is to estimate $x$ and round the estimate. Flip every coin $m$ times: the
$t$-th round of flips is a vector $X_t\in\{0,1\}^{d}$ with $\E X_t=x$, and the empirical mean
\[
   Z=\frac1m\sum_{t=1}^{m}X_t
\]
is an unbiased estimate of $x$. It \revcolor{need not be} a point of $P$, and it need not even lie on
$\aff(P)$, so a vertex decomposition of $Z$ makes no sense. Projecting onto $\aff(P)$ repairs
this at no cost: put $Y=\Pi(Z)$. Since $\Pi$ is affine and fixes $\aff(P)$ pointwise,
\begin{equation}\label{eq:EY}
   \E Y=\Pi(\E Z)=\Pi(x)=x\qquad\text{exactly, for every }m .
\end{equation}
The projection may violate feasibility: $Y$ lies on $\aff(P)$ but not necessarily in $P$
(Figure~\ref{fig:batch}). The promise makes this rare. The ball
$\mathbb B(x,\rho)\cap\aff(P)$ lies inside $P$, and by Hoeffding's inequality every coordinate of
$Z$ is within $\rho/\sqrt d$ of the corresponding coordinate of $x$, hence
$\|Z-x\|_2\le\rho$, except with probability $2de^{-2m\rho^{2}/d}$. Projection does not increase
the distance to a point of $\aff(P)$, so on that event $Y\in\mathbb B(x,\rho)\cap\aff(P)\subseteq
P$. Once $m$ exceeds a constant times $(d/\rho^{2})\log d$ this probability is below any fixed
inverse polynomial in $d$, and it keeps decreasing exponentially in $m$.

So the simplest scheme is: if $Y\in P$ output $\Round(Y)$, and otherwise output $\Round(u)$.
Every output is a vertex, and \revcolor{by \eqref{eq:EY},} the mean of the point handed to $\Round$ is
\[
   \E\bigl[Y\,\one\{Y\in P\}+u\,\one\{Y\notin P\}\bigr]
   =\E Y-\E\bigl[(Y-u)\,\one\{Y\notin P\}\bigr]
   =x-\E\bigl[(Y-u)\,\one\{Y\notin P\}\bigr],
\]
\revcolor{where} the subtracted term has $\ell_2$ norm at most $\sqrt d\;\Prb[Y\notin P]$ \revcolor{because} $\|Y-u\|_2\le\|Z-u\|_2\le\sqrt d$. This \revcolor{scheme} is feasible and almost exact: the
error is exponentially small in $m$\revcolor{, but} in general it is not zero for any $m$.
The remaining work is to compensate for the event $Y\notin P$.

\subsection{A reserve}\label{sec:reserve}

The scheme above needs a small fix before any correction can be attached to it. Suppose that
with probability $1-\alpha$ we round $Y$ and with probability $\alpha$ we run some correction. The
correction will turn out to be a small perturbation of $u$, so the mean of the output is close to
$(1-\alpha)x+\alpha u$ rather than to $x$: rounding $Y$ itself on the main branch leaves no room
for the correction. The fix is geometric. Expand $Y$ away from $u$ by the factor $1/(1-\alpha)$,
\[
   \frac{Y-\alpha u}{1-\alpha}=u+\frac{Y-u}{1-\alpha},
\]
so that mixing the expanded point with $u$ in proportions $1-\alpha:\alpha$ gives back $Y$.
Rounding the expanded point with probability $1-\alpha$ and a point near $u$ with probability
$\alpha$ then has mean near $\E Y=x$, and the correction only has to account for a small error.
Define
\begin{equation}\label{eq:Ytilde}
   \tilde Y=\begin{cases}
     \dfrac{Y-\alpha u}{1-\alpha} & \text{if this point lies in }P,\\[8pt]
     u & \text{otherwise.}
   \end{cases}
\end{equation}
One membership test decides which case occurs, and in both cases $\tilde Y\in P$. Geometrically
the expansion changes little. We take $\alpha=\rho/(4s)$, so that $\alpha\|x-u\|_2\le\rho/4$
because $\|x-u\|_2\le\sqrt d\le s$. If $\|Z-x\|_\infty\le\rho/(4s)$ then $\|Y-x\|_2\le\rho/4$
and
\[
   \Bigl\|\frac{Y-\alpha u}{1-\alpha}-x\Bigr\|_2
   =\frac{\|(Y-x)+\alpha(x-u)\|_2}{1-\alpha}
   \le\frac{\rho/4+\rho/4}{3/4}=\frac{2\rho}{3},
\]
so the expanded point lies in $\mathbb B(x,\rho)\cap\aff(P)\subseteq P$ and the first case of
\eqref{eq:Ytilde} occurs. By Hoeffding's inequality the second case has probability at most
$2de^{-m\rho^{2}/(8s^{2})}$.

Now we compute $\E\tilde Y$. Write $R$ for the \emph{residual}, $R=Y-u$ in the second case of
\eqref{eq:Ytilde} and $R=0$ in the first, and $\bar R=\E R$ for its mean; then
$\|\bar R\|_2\le\sqrt d\,\Prb[R\ne0]$, and in both cases
\[
   (1-\alpha)\tilde Y=Y-\alpha u-R .
\]
Taking expectations and using \eqref{eq:EY},
\begin{equation}\label{eq:mainbranch}
   (1-\alpha)\,\E\tilde Y=x-\alpha u-\bar R .
\end{equation}
The plan is now the following. With probability $1-\alpha$ we output $\Round(\tilde Y)$, which by
\eqref{eq:mainbranch} contributes $x-\alpha u-\bar R$ to the mean of the output; with probability
$\alpha$ we run a correction, which must therefore output a vertex whose mean is
\begin{equation}\label{eq:targetpoint}
   u+\frac{\bar R}{\alpha}.
\end{equation}
\revcolor{We know two facts about this point: (i) it is very close to $u$, within distance
$\sqrt d\,\Prb[R\ne0]/\alpha$, so it lies in $\mathbb B(u,r)\cap\aff(P)\subseteq P$ as soon as the
batch fails with probability at most $\alpha r/\sqrt d$; and (ii) its displacement from $u$ is
unknown, since $\bar R$ is the expectation of a quantity that the algorithm observes, not a number
that it can compute. Next, we design the correction.}

\subsection{The correction: a lottery over anchors}\label{sec:anchors}

\begin{figure}[t]
\centering
\begin{tikzpicture}[scale=1.0,>=latex]
  \draw[fill=black!4,draw=black!45] (0,0) -- (4.3,0.45) -- (5.05,2.6) -- (3.1,4.25) -- (0.7,3.55) -- (-0.55,1.6) -- cycle;
  \node[black!55,anchor=west,inner sep=2pt] at (4.2,3.6) {$P$};
  \coordinate (u) at (2.2,2.1);
  \draw[dashed,black!55] (u) circle (1.25);
  \draw[<->,black!60] (u) -- ($(u)+(37:1.25)$);
  \node[black!70,inner sep=1pt] at ($(u)+(37:0.72)$) [above left=-2pt] {$r$};
  \foreach \a/\lab/\pos in {0/{$u_1^{+}$}/right, 180/{$u_1^{-}$}/left,
                            90/{$u_2^{+}$}/above, 270/{$u_2^{-}$}/below}
    { \fill ($(u)+(\a:1.25)$) circle (1.8pt);
      \node[\pos,inner sep=3pt] at ($(u)+(\a:1.25)$) {\lab}; }
  \coordinate (t) at (2.85,1.70);
  \draw[->,black!65,shorten >=2pt] (u) -- (t);
  \fill (t) circle (1.8pt);
  \node[anchor=north east,inner sep=3pt] at (t) {$u+\bar R/\alpha$};
  \fill (u) circle (1.8pt); \node[anchor=south east,inner sep=2pt] at (u) {$u$};
\end{tikzpicture}
\caption{The anchors, drawn for $\dim\aff(P)=2$. The ball $\mathbb B(u,r)\cap\aff(P)$ lies in $P$,
so the $2d$ points $u_i^{\pm}=\Pi(u\pm re_i)$ are feasible. The correction picks $i$ uniformly,
chooses between $u_i^{+}$ and $u_i^{-}$ with a coin tilted by the observed residual, and
decomposes the chosen point into vertices; the resulting vertex lottery has mean exactly
$u+\bar R/\alpha$, a point inside the ball that the algorithm never computes.}
\label{fig:probe}
\end{figure}
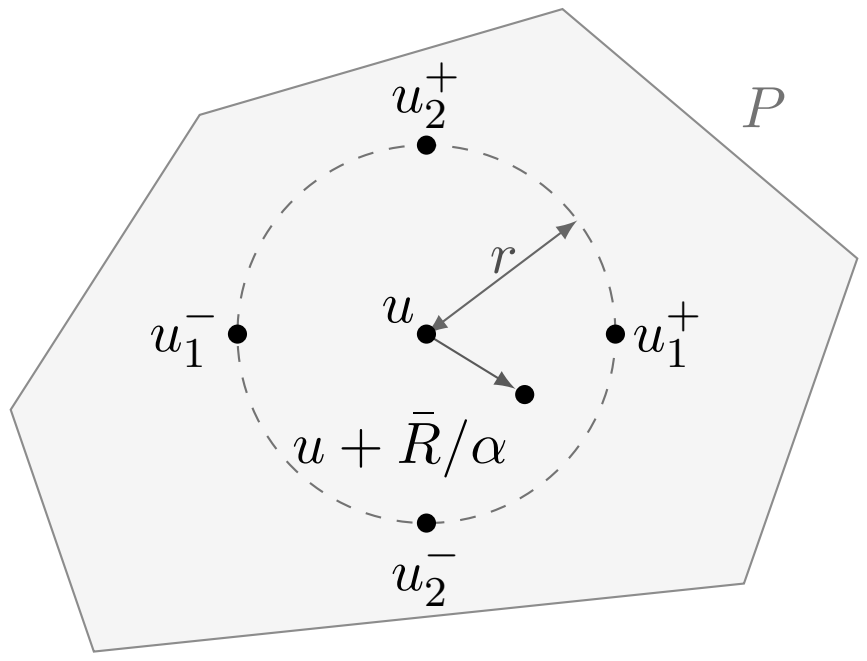

The $2d$ \emph{anchors}
\[
   u_i^{\pm}=\Pi(u\pm r\,e_i)=u\pm r\,\Pi_L e_i\qquad(i\in[d])
\]
are feasible, because $\|\Pi_L e_i\|_2\le1$ puts them in $\mathbb B(u,r)\cap\aff(P)$, and the
midpoint of every pair is $u$ (Figure~\ref{fig:probe}). A lottery that picks $i$ uniformly and
then $u_i^{+}$ with probability $h_i$ and $u_i^{-}$ with probability $1-h_i$ has mean
\[
   u+\frac rd\sum_{i=1}^{d}(2h_i-1)\,\Pi_L e_i .
\]
Taking $2h_i-1=\frac{d}{r\alpha}\bar R_i$ makes this $u+\frac1\alpha\Pi_L \bar R=u+\bar R/\alpha$, because $\bar R\in
L$ is fixed by $\Pi_L$. \revcolor{So this lottery reaches the target \eqref{eq:targetpoint}, and since
it is a lottery over points of $P$, the feasibility of the target needs no separate argument. The
requirement $|2h_i-1|\le1$ amounts to $\|\bar R\|_\infty\le r\alpha/d$, which again asks for a
small residual.}

It remains to flip a coin of bias $h_i=\frac12+\frac{d}{2r\alpha}\bar R_i$ without knowing $\bar R_i$. A
fresh batch produces a sample of $R$, and $|R_i|\le\|R\|_2\le\sqrt d\le s$, so
\[
   p_i^{+}=\E\Bigl[\frac{\pos{R_i}}{s}\Bigr]
   \qquad\text{and}\qquad
   p_i^{-}=\E\Bigl[\frac{\pos{-R_i}}{s}\Bigr]
\]
are biases of coins the algorithm can flip: run a batch, compute $R$, and flip a coin with the
observed rational probability $\pos{\pm R_i}/s$. Their difference is $\bar R_i/s$, so
\[
   h_i=\frac12\bigl(1+\beta\,p_i^{+}-\beta\,p_i^{-}\bigr),
   \qquad \beta=\frac{ds}{r\alpha},
\]
and a coin of bias $h_i$ is a fair mixture of a $\beta p_i^{+}$-coin and the complement of a
$\beta p_i^{-}$-coin. Turning a $p$-coin into a $\beta p$-coin is the linear Bernoulli factory of
Definition~\ref{def:linear}: it is exact, it needs the promise $\beta p\le\tfrac14$, and it costs
at most $K_0\beta$ flips of the $p$-coin \revcolor{in expectation}. The promise holds because $p_i^{\pm}\le\Prb[R\ne0]$ and
$m$ is chosen so that $\Prb[R\ne0]\le1/(4\beta)$. Nothing here estimates $\bar R$: the factory turns a
coin of unknown bias $p_i^{\pm}$ into a coin of bias exactly $\beta p_i^{\pm}$, and the lottery
turns that into a vertex whose mean is exactly $u+\bar R/\alpha$.

The correction is expensive: one call to the linear factory consumes up to $K_0\beta$
batches \revcolor{in expectation}, and
$\beta=4ds^{2}/(r\rho)$ is large. But it is run only with probability $\alpha=\rho/(4s)$, and
\[
   \alpha\beta=\frac{ds}{r}
\]
does not depend on $\rho$ at all. \revcolor{So the expected number of batches spent on the correction is at most
$\alpha\cdot K_0\beta=K_0ds/r$.}

\subsection{The algorithm}\label{sec:algorithm}

With $s=\lceil\sqrt d\rceil$ the parameters are
\begin{equation}\label{eq:idea-params}
   \alpha=\frac{\rho}{4s},\qquad
   \beta=\frac{ds}{r\alpha}=\frac{4ds^{2}}{r\rho},\qquad
   m=\Bigl\lceil\frac{8s^{2}}{\rho^{2}}\Bigr\rceil\bigl\lceil\log_{2}(8d\beta)\bigr\rceil ,
\end{equation}
all rational and computed exactly; $\lceil\log_2 z\rceil$ is found by comparing $z$ with powers of
two.

\begin{procedure}{$\Fact$}
\begin{enumerate}[leftmargin=2em,itemsep=2pt,topsep=2pt]
\item Flip a coin of bias $\alpha$. On tails go to step 2; on heads go to step 3.
\item \emph{(Main branch.)} Flip each input coin $m$ times, let $Z$ be the vector of empirical
  means and $Y=\Pi(Z)$, and form $\tilde Y$ as in \eqref{eq:Ytilde} with one call to $\Mem$.
  Return $\Round(\tilde Y)$.
\item \emph{(Correction.)} Draw $i$ uniformly from $[d]$ and a uniform sign
  $\sigma\in\{+1,-1\}$. Run the linear factory $\Lin(\beta,\cdot)$, and whenever it asks for a
  coin flip, produce one as follows: flip each input coin $m$ fresh times, form $Y$ and
  $\tilde Y$ as in step 2, let $R=Y-u$ if the membership test failed and $R=0$ otherwise, and
  return heads with probability $\pos{\sigma R_i}/s$. Let $\xi$ be the factory's output if
  $\sigma=+1$ and its complement if $\sigma=-1$. Return $\Round(u_i^{+})$ if $\xi=1$ and
  $\Round(u_i^{-})$ otherwise.
\end{enumerate}
\end{procedure}

\begin{theorem}[Interior exact rounding]\label{thm:interior}
Let $u\in\Int_r(P)$ be a rational point and $x\in\Int_\rho(P)$ the unknown coin vector, with
rational $r,\rho\in(0,1]$. Then $\Fact$ terminates almost surely and outputs $V\in\vertx(P)$ with
$\E V=x$ exactly. It makes one call to $\Round$, at most $1+K_0ds/r$ calls to $\Mem$ in
expectation, and at most
\[
   \Bigl(1+K_0\frac{ds}{r}\Bigr)\,d\,m
   \ =\ O\!\left(\frac{d^{7/2}}{r\,\rho^{2}}\log\frac{d}{r\rho}\right)
\]
input-coin flips in expectation. Its expected running time is polynomial in $d$, $1/r$, $1/\rho$,
the encoding lengths of $u$, $r$, $\rho$ and the description of $\aff(P)$, and the stipulated cost
of $\Mem$ and $\Round$.
\end{theorem}

\subsection{Proof of Theorem~\ref{thm:interior}}\label{sec:proofinterior}

The proof collects the facts established along the way, in two lemmas: one about a single batch,
one about the correction.

\begin{lemma}[The batch]\label{lem:batch}
One batch of $m$ rounds, with $\alpha=\rho/(4s)$, produces rational $\tilde Y$ and $R$ with
\begin{align}
 &\tilde Y\in P; \label{eq:feas}\\
 &R\in L,\qquad (1-\alpha)\tilde Y=Y-\alpha u-R,\qquad (1-\alpha)\,\E\tilde Y=x-\alpha u-\bar R;
   \label{eq:unb}\\
 &|R_i|\le\|R\|_2\le\sqrt d\quad(i\in[d]); \label{eq:bdd}\\
 &\Prb[R\ne0]\ \le\ 2d\,e^{-m\rho^{2}/(8s^{2})}\ \le\ \frac1{4\beta}. \label{eq:rare}
\end{align}
It uses $dm$ input-coin flips and one call to $\Mem$, and every rational number it forms has
encoding length polynomial in $\log m$, $\log d$ and the encoding lengths of $u$, $\alpha$ and
$\aff(P)$.
\end{lemma}

\begin{proof}
\eqref{eq:feas}: in the first case of \eqref{eq:Ytilde} the point was accepted by $\Mem$, and in
the second $\tilde Y=u\in P$.

\eqref{eq:unb}: the pointwise identity was checked in both cases in Section~\ref{sec:reserve};
$R\in\{0,Y-u\}$ and $Y-u=\Pi(Z)-\Pi(u)=\Pi_L(Z-u)\in L$; taking expectations and using
\eqref{eq:EY} gives the last identity.

\eqref{eq:bdd}: $\|R\|_2\le\|\Pi_L(Z-u)\|_2\le\|Z-u\|_2\le\sqrt d$, because $\Pi_L$ is an
orthogonal projection and $Z,u\in[0,1]^{d}$.

\eqref{eq:rare}: suppose $\|Z-x\|_\infty\le\rho/(4s)$. Then
$\|Y-x\|_2=\|\Pi_L(Z-x)\|_2\le\|Z-x\|_2\le\sqrt d\,\|Z-x\|_\infty\le\rho/4$, and the computation
in Section~\ref{sec:reserve} places $(Y-\alpha u)/(1-\alpha)$ in $\mathbb B(x,2\rho/3)\cap\aff(P)
\subseteq P$, so the first case of \eqref{eq:Ytilde} occurs and $R=0$. Hoeffding's inequality
\cite{Hoeffding1963} for each of the $d$ coordinates of $Z$ and a union bound give
\[
   \Prb[R\ne0]\ \le\ \Prb\bigl[\|Z-x\|_\infty>\rho/(4s)\bigr]
   \ \le\ 2d\exp\Bigl(-\frac{2m\rho^{2}}{16s^{2}}\Bigr),
\]
and $m\ge\frac{8s^{2}}{\rho^{2}}\log_2(8d\beta)\ge\frac{8s^{2}}{\rho^{2}}\ln(8d\beta)$ makes this at
most $2d/(8d\beta)=1/(4\beta)$.

The flip and oracle counts are immediate. $Z$ has denominator $m$, so $Y$, $\tilde Y$ and $R$ are
obtained from $Z$, $u$, $\alpha$ and the rational description of $\Pi$ by a bounded number of
rational operations, which gives the encoding bound.
\end{proof}

\begin{lemma}[The correction]\label{lem:correct}
Step 3 of $\Fact$ outputs a vertex $C$ of $P$ with
\[
   \E C=u+\frac{\bar R}{\alpha},
\]
and uses at most $K_0\beta$ batches in expectation.
\end{lemma}

\begin{proof}
Fix $i$ and $\sigma$. A coin flip requested by the linear factory is a fresh batch followed by a
coin of bias $\pos{\sigma R_i}/s$, a rational number in $[0,1]$ by \eqref{eq:bdd}; so the factory
is fed independent coins of bias $p_i^{\sigma}=\E[\pos{\sigma R_i}/s]$. Since
$\pos{\sigma R_i}/s\le\one\{R\ne0\}$ pointwise, \eqref{eq:rare} gives
$\beta p_i^{\sigma}\le\beta\,\Prb[R\ne0]\le\tfrac14$, so the factory is used within its promise:
it returns a bit of mean exactly $\beta p_i^{\sigma}$ after at most $K_0\beta$ coin flips \revcolor{(that is, batches) in expectation}. Averaging over the sign,
\[
   \Prb[\xi=1\mid i]
   =\tfrac12\bigl(\beta p_i^{+}+1-\beta p_i^{-}\bigr)
   =\tfrac12+\tfrac\beta2\bigl(p_i^{+}-p_i^{-}\bigr)
   =\tfrac12+\frac{\beta}{2s}\bar R_i ,
\]
because $\pos t-\pos{-t}=t$ gives $p_i^{+}-p_i^{-}=\E R_i/s=\bar R_i/s$; this probability lies in
$[\tfrac38,\tfrac58]$. Conditionally on $i$, the point handed to $\Round$ is $u_i^{+}$ with this
probability and $u_i^{-}$ otherwise, so its conditional mean is
$u+(2\Prb[\xi=1\mid i]-1)\,r\,\Pi_L e_i=u+\frac{\beta r}{s}\bar R_i\,\Pi_L e_i$. Averaging over the
uniform $i$, using $\beta r/(ds)=1/\alpha$ and $\Pi_L \bar R=\bar R$ (as $\bar R\in L$ by \eqref{eq:unb}),
\[
   \E\bigl[\text{point handed to }\Round\bigr]
   =u+\frac{\beta r}{ds}\sum_{i=1}^{d}\bar R_i\,\Pi_L e_i
   =u+\frac1\alpha\,\Pi_L \bar R=u+\frac{\bar R}{\alpha} .
\]
The points $u_i^{\pm}=u\pm r\Pi_L e_i$ are rational and lie in $\mathbb B(u,r)\cap\aff(P)\subseteq
P$ because $\|r\Pi_L e_i\|_2\le r$ and $u\in\Int_r(P)$; so $\Round$ returns a vertex, and \revcolor{since $\Round$ is mean
preserving, $\E C=u+\bar R/\alpha$}.
\end{proof}

\begin{proof}[Proof of Theorem~\ref{thm:interior}]
\emph{Vertices.} Every point handed to $\Round$ is a rational point of $P$: $\tilde Y$ in step 2
by \eqref{eq:feas}, and $u_i^{\pm}$ in step 3. So the output is a vertex of $P$.

\emph{Exactness.} Step 2 outputs a vertex of mean $\E\tilde Y$, step 3 a vertex of mean
$u+\bar R/\alpha$ by Lemma~\ref{lem:correct}, and the branch is chosen by an independent
$\alpha$-coin. By \eqref{eq:unb},
\[
   \E V=(1-\alpha)\,\E\tilde Y+\alpha\Bigl(u+\frac{\bar R}{\alpha}\Bigr)
       =(x-\alpha u-\bar R)+\alpha u+\bar R=x .
\]
Nothing is discarded: step 2 loses the residual $R$ and step 3 restores it. The concentration
bound \eqref{eq:rare} was used only to certify the promise of the linear factory; it never enters
the computation of $\E V$.

\emph{Cost and termination.} Step 2 uses one batch. Step 3 uses at most $K_0\beta$ batches in
expectation by Lemma~\ref{lem:correct} and is entered with probability $\alpha$; since
$\alpha\beta=ds/r$, the expected number of batches is at most
$(1-\alpha)+\alpha K_0\beta\le1+K_0ds/r$. Each batch costs $dm$ input flips and one call to
$\Mem$, and each branch ends with one call to $\Round$, which gives the flip and oracle bounds.
For the asymptotic form use $s\le2\sqrt d$, $\beta\le16d^{2}/(r\rho)$ and
$m=O\bigl(\frac{d}{\rho^{2}}\log\frac{d}{r\rho}\bigr)$. The expected number of batches is finite
and the linear factory terminates almost surely (Appendix~\ref{app:scalar}), so $\Fact$
terminates almost surely.

\emph{Bit complexity.} By Lemma~\ref{lem:batch} every rational the algorithm forms has encoding
length polynomial in the encoding lengths of $u$, $r$, $\rho$ and $\aff(P)$, in $\log d$ and in
$\log m=\revcolor{O(\log(d/(r\rho)))}$; each batch is fresh, so the state of the linear factory does not
inflate these lengths. The linear factory has expected bit cost polynomial in $\beta$ and
$\bits(\beta)$ (Appendix~\ref{app:scalar}), and every known rational coin is flipped exactly from
fair bits at expected cost polynomial in its encoding length. Combining these with the stipulated
cost of $\Mem$ and $\Round$ and the expected number of calls gives the running-time bound.
\end{proof}

\section{Universal factories on affine slices}\label{sec:nomargin}

When $P$ is an affine slice of the cube, $P=\aff(P)\cap[0,1]^{d}$, we can do better than
Theorem~\ref{thm:interior}: we give an algorithm that is correct at every point of the relative
interior $\Int(P)$ and is told nothing about $x$. These polytopes are special\revcolor{:} Niazadeh, Paes
Leme\revcolor{,} and Schneider \cite{NPS2023} showed that they are exactly the polytopes that admit a
Bernoulli factory, that is, an algorithm correct at every point of $P\cap(0,1)^{d}$ with no
further promise on $x$ \revcolor{(for an affine slice with no constant coordinate, this domain is
$\Int(P)$). For generic slices, their factory terminates at every point of $P$, including the
boundary}. Our factory does not terminate on the boundary of $P$. In exchange, at every
$x\in\Int(P)$\revcolor{,} its expected number of coin flips is polynomial in $d$ and $1/\dmin$, where
$\dmin=\min_i\{x_i,1-x_i\}$ is the $\ell_\infty$ distance from $x$ to the boundary of the cube.
The guarantee depends on $\dmin$, but the algorithm is never given \revcolor{$\dmin$}.

\paragraph{The connection with the characterization.} \revcolor{Both the characterization of
\cite{NPS2023} and the design of our algorithm rest} on one property of factories. A leaf is reached with probability a Bernstein monomial \eqref{eq:bernstein}, which is
either identically zero or positive on all of $(0,1)^{d}$; so the set of transcripts with
positive probability depends only on which coordinates of $x$ equal $0$ or $1$, and is the same
at every point of $(0,1)^{d}$. \revcolor{Niazadeh et al.}~\cite{NPS2023} use this
property to prove that a factory exists only if no proper face $F$ of $P$ meets the open cube:
otherwise a point of $F\cap(0,1)^{d}$ and a point of $\Int(P)$ reach the same leaves with
positive probability, yet the first must output only vertices of $F$ and the second must output
some vertex outside $F$. An affine slice has no such face, because a point of $P\cap(0,1)^{d}$ has
a neighborhood in $\aff(P)$ that lies in the cube and hence in $P$. Lemma~\ref{lem:cuberadius}
below makes this quantitative: the neighborhood has radius $\dmin$. \revcolor{So the margin that had to be given to the algorithm in Section~\ref{sec:idea}} is
now a statistic of the coins, and a pilot can estimate it. The same property of Bernstein monomials prevents the estimate from ever being certified,
since a transcript with positive probability at $x$ has positive probability at points of
$(0,1)^{d}$ with arbitrarily small $\dmin$. Our algorithm therefore uses the estimate only through
a guarantee that holds on average.

\medskip
Throughout this section\revcolor{,} $P$ is such a slice, given by a rational system $Wz=b$ describing
$\aff(P)$ and with the oracles $\Mem$ and $\Round$ of Section~\ref{sec:prelim}\revcolor{. We assume
that no coordinate is constant on $P$, so that $\Int(P)=P\cap(0,1)^{d}$, and that the unknown $x$
lies in $\Int(P)$. As before, $s=\lceil\sqrt d\rceil$.} Besides the oracles, the algorithm of Section~\ref{sec:idea} is
given a reference point $u\in\Int_r(P)$ with its radius $r$, and a radius $\rho$ with
$x\in\Int_\rho(P)$, which sets the reserve $\alpha$ and the batch size $m$. On an affine slice\revcolor{,} $\dmin$ takes the place of $\rho$ \revcolor{(Lemma~\ref{lem:cuberadius})},
and the reference point comes from a linear program \revcolor{(Lemma~\ref{lem:lpcenter})}.

\begin{lemma}\label{lem:cuberadius}
If $P=\aff(P)\cap[0,1]^{d}$ and $x\in P$, then $x\in\Int_t(P)$ for every $t\le\dmin$.
\end{lemma}

\begin{proof}
Let $z\in\mathbb B(x,t)\cap\aff(P)$. Then $\|z-x\|_\infty\le\|z-x\|_2\le t\le\dmin$, so
$z\in[0,1]^{d}$, and $z\in\aff(P)$; hence $z\in P$.
\end{proof}

\begin{lemma}[A reference point by linear programming]\label{lem:lpcenter}
Let $(u,r)$ be an optimal solution of the linear program
\[
   \max\ t\quad\text{subject to}\quad Wz=b,\qquad t\le z_{i}\le1-t\ \ (i\in[d]),
\]
which is feasible and has optimum at most $\tfrac12$. Then $r>0$, $u\in\Int_r(P)$, and
$r\ge\dmin$ for every $x\in\Int(P)$. The program has polynomial size, so $u$ and $r$ are rational
and computed exactly in polynomial time.
\end{lemma}

\begin{proof}
The pair $(x,\dmin)$ is feasible for every $x\in\Int(P)$, so $r\ge\dmin>0$. If
$z\in\mathbb B(u,r)\cap\aff(P)$ then $\|z-u\|_\infty\le r$, so $z\in[0,1]^{d}$ because
$r\le u_i\le1-r$ for every $i$; hence $z\in P$, and $u\in\Int_r(P)$.
\end{proof}

So $u$ and $r$ are known, and the only missing input is a lower bound on $\dmin$.

\subsection{The idea}\label{sec:univ-idea}

\revcolor{We want a factory for $P$ that is correct on an open domain with no parameter supplied, at a
cost that depends on where the unknown bias lies; this is the setting of Keane and O'Brien's
characterization of scalar factories \cite{KeaneOBrien1994}.} In Section~\ref{sec:idea}\revcolor{,} a
wrong value of $\rho$ would break only one step, \revcolor{namely} the linear factory
$p\mapsto\beta p$, which
must be told its slack \cite{NacuPeres2005,Huber2016}; every other step is exact for every choice
of $\alpha$ and $m$. To remove $\rho$ we borrow two techniques. The first is Morina's view of a Bernoulli factory as an
unbiased $[0,1]$-valued estimator of $f(p)$, constructed by debiasing a biased one
\cite{Morina2021} and developed further by Koskela, Karvonen, \L{}atuszy\'nski\revcolor{,} and Span\`o
\cite{Koskela2025}. Section~\ref{sec:idea} already follows this view for a $P$-valued estimator of
$x$: the batch is the biased estimator, and the correction is the debiasing step, realized as a
lottery over feasible points so that the estimator stays in $P$. The second is the two-coin
algorithm of Gon\c{c}alves, \L{}atuszy\'nski\revcolor{,} and Roberts \cite{GLR2017,GLR2023}, in the form of
the division factory of Nacu and Peres \cite{NacuPeres2005}, which we use to divide by an unknown
mean. Those algorithms divide by a coin whose gap to the numerator is known; we \revcolor{observe}
that no lower bound on the denominator is needed \revcolor{when the division is triggered by an event
whose probability is the denominator itself, since the reciprocal cost of the division then
cancels}. The pilot, its averaged guarantee\revcolor{,} and the random reserve are, to our knowledge,
new. The algorithm has three parts.

\paragraph{A pilot.} Consider the ladder of scales $\lambda_\ell=2^{-\ell-1}$, $\ell=0,1,2,\dots$\revcolor{.}
At stage $\ell$ the pilot flips every coin $N_\ell$ times and passes if every coin shows at least
$h_\ell$ heads and at least $h_\ell$ tails; it returns the scale $\lambda=\lambda_\ell$ of the first
passing stage. The counts are chosen so that a stage with $\lambda_\ell>\dmin$ passes with
probability at most $\psi\dmin^{2}/\lambda_\ell$, for a small known $\psi$, and a stage with
$\lambda_\ell\le\dmin/8$ fails with probability at most $\tfrac1{64}$. So $\lambda$ typically lies
between $\dmin/16$ and $\dmin$, where it is a valid radius at $x$ by Lemma~\ref{lem:cuberadius}.
\revcolor{It can also overshoot, and the algorithm cannot detect this.} Summing the overshoot bounds, and using
$\E[\lambda]\ge\dmin/32$, gives an \emph{averaged} guarantee instead,
\begin{equation}\label{eq:averaged-idea}
   \Prb[\lambda>\dmin]\ \le\ 2\psi\dmin\ \le\ 64\psi\,\E[\lambda]\revcolor{,}
\end{equation}
\revcolor{which says that} the pilot's mistakes are rare in proportion to the size of the reserve
\revcolor{that pays for them}.

\paragraph{A random reserve.} Given $\lambda$, we run the batch of Section~\ref{sec:idea} with
$\lambda$ in place of $\rho$: reserve $\widehat\alpha=\lambda/(4s)$, batch size $m(\lambda)$, and
$\tilde Y$ defined by \eqref{eq:Ytilde} with $\widehat\alpha$ for $\alpha$. The reserve is now
random and correlated with the batch, but the pointwise identity
$(1-\widehat\alpha)\tilde Y=Y-\widehat\alpha u-R$ of Section~\ref{sec:reserve} holds for every
value of $\lambda$, and taking expectations gives
\[
   \E\bigl[(1-\widehat\alpha)\tilde Y\bigr]=x-a\,u-\bar R,
   \qquad a=\E[\widehat\alpha],\quad \bar R=\E[R].
\]
The algorithm flips its $\widehat\alpha$-coin after the batch, rounds $\tilde Y$ on tails and
corrects on heads, so it enters the correction with probability exactly $a$. As in
\eqref{eq:targetpoint}, the correction must therefore output a vertex of mean
\begin{equation}\label{eq:targetpoint-univ}
   u+\frac{\bar R}{a},
\end{equation}
with the residual divided by the \emph{mean} reserve $a$\revcolor{, which is unknown, and not by the
observed $\widehat\alpha$, which the algorithm knows but cannot trust}.

\paragraph{Dividing by an unknown mean.} In Section~\ref{sec:anchors} the coin choosing between the
anchors $u_i^{\pm}$ was tilted by $\beta p_i^{\pm}$ with $\beta=c/\alpha$, where $c=ds/r$. Now the
tilt is $c\,p_i^{\pm}/a$, a ratio of two unknown means. A fresh pilot and batch followed by a coin
of bias $\widehat\alpha$ is a coin of mean exactly $a$. Race it against the $p$-coin with odds
$c:1$, as in the two-coin algorithm of \cite{GLR2017}: the first head comes from the $p$-coin with
probability $q=cp/(a+cp)$, and
\[
   \frac{q}{1-q}=\frac{cp}{a}=\sum_{j\ge1}2^{-j}(2q)^{j},
\]
so a geometric number of $2q$-coins, each made from $q$-coins by the linear factory
$\Lin(2,\cdot)$, yields a coin of bias exactly $cp/a$. The race costs \revcolor{at most} $(c+1)/a$ source calls in
expectation and uses no lower bound on $a$. The factor $1/a$ is paid only on the correction branch,
which is entered with probability $a$, so the two cancel as $\alpha$ and $\beta$ did in
Section~\ref{sec:anchors}, and the expected cost is $O(c)$ batches. The ratio factory needs $cp/a$
to stay below a constant, which holds if $c\,\Prb[R\ne0]=O(a)$. On $\{\lambda\le\dmin\}$ the radius
$\lambda$ is valid, so by the argument of Section~\ref{sec:reserve} a batch of size $m(\lambda)$
fails with probability $O(\widehat\alpha/c)$, which averages to $O(a/c)$; on
$\{\lambda>\dmin\}$ we assert nothing, and \eqref{eq:averaged-idea} bounds its probability by
$O(a/c)$ as well.

\subsection{The algorithm}\label{sec:univ-alg}

With $s=\lceil\sqrt d\rceil$ and $(u,r)$ from Lemma~\ref{lem:lpcenter}, the parameters are
\begin{equation}\label{eq:univ-params}
   c=\frac{ds}{r},\qquad
   \psi=\frac1{2^{14}\,d\,s\,c},\qquad
   \lambda_{\ell}=2^{-\ell-1},\qquad
   h_{\ell}=4\Bigl\lceil\log_{2}\frac1{\psi\lambda_{\ell}}\Bigr\rceil,\qquad
   N_{\ell}=\frac{h_{\ell}}{4\lambda_{\ell}},
\end{equation}
and, for a scale $\lambda$ returned by the pilot,
\begin{equation}\label{eq:univ-params2}
   \widehat\alpha(\lambda)=\frac{\lambda}{4s},\qquad
   m(\lambda)=\Bigl\lceil\frac{8s^{2}}{\lambda^{2}}\Bigr\rceil
              \Bigl\lceil\log_{2}\frac{512\,d\,s\,c}{\lambda}\Bigr\rceil .
\end{equation}
Every $N_\ell$ is an integer, $h_\ell=h_0+4\ell$, and all quantities are rational and computed
exactly. None of them depends on $x$.

\begin{procedure}{$\Univ$}
An \emph{adaptive batch} is the following two-part procedure; each call uses fresh coin flips.
\begin{enumerate}[leftmargin=2em,itemsep=2pt,topsep=2pt]
\item[(i)] \emph{Pilot.} For $\ell=0,1,2,\dots$: flip each input coin $N_\ell$ times; if every coin
  shows at least $h_\ell$ heads and at least $h_\ell$ tails, set $\lambda=\lambda_\ell$ and stop.
\item[(ii)] \emph{Batch.} Flip each input coin $m(\lambda)$ times, let $Z$ be the vector of
  empirical means and $Y=\Pi(Z)$, and form $\tilde Y$ as in \eqref{eq:Ytilde} with
  $\widehat\alpha=\widehat\alpha(\lambda)$ in place of $\alpha$, using one call to $\Mem$; let
  $R=Y-u$ if the membership test failed and $R=0$ otherwise. Return $(\lambda,\tilde Y,R)$.
\end{enumerate}
The algorithm itself:
\begin{enumerate}[leftmargin=2em,itemsep=2pt,topsep=2pt]
\item Run an adaptive batch to obtain $(\lambda,\tilde Y,R)$, and flip a coin of bias
  $\widehat\alpha(\lambda)$. On tails return $\Round(\tilde Y)$. On heads discard
  $(\lambda,\tilde Y,R)$ and continue.
\item \emph{(Correction.)} Draw $i$ uniformly from $[d]$ and a uniform sign
  $\sigma\in\{+1,-1\}$. Produce a bit $\xi'$ as follows.
  \begin{enumerate}[label=(\alph*),leftmargin=1.6em,itemsep=1pt,topsep=1pt]
  \item A \emph{$q$-coin} is flipped by repeating the following until a head occurs: with
    probability $c/(c+1)$ run an adaptive batch and flip a coin of bias $\pos{\sigma R_i}/s$
    from its $R$; with probability $1/(c+1)$ run an adaptive batch and flip a coin of bias
    $\widehat\alpha(\lambda)$ from its $\lambda$. The $q$-coin shows $1$ if the first head came
    from a coin of the first kind and $0$ otherwise.
  \item Draw $G$ with $\Prb[G=j]=2^{-j}$ for $j\ge1$. Run the linear factory $\Lin(2,\cdot)$ on
    independent $q$-coins up to $G$ times, stopping at the first output $0$; set $\xi'=1$ if all
    $G$ runs output $1$ and $\xi'=0$ otherwise.
  \end{enumerate}
  Let $\xi=\xi'$ if $\sigma=+1$ and $\xi=1-\xi'$ if $\sigma=-1$. Return $\Round(u_i^{+})$ if
  $\xi=1$ and $\Round(u_i^{-})$ otherwise, where $u_i^{\pm}=\Pi(u\pm re_i)$ are the anchors of
  Section~\ref{sec:anchors}.
\end{enumerate}
\end{procedure}

\begin{theorem}[Universal factories on affine slices]\label{thm:universal}
Let $P=\aff(P)\cap[0,1]^{d}$ be a rational affine slice of the cube with
$\Int(P)\ne\varnothing$. For every $x\in\Int(P)$, $\Univ$ terminates almost surely and outputs
$V\in\vertx(P)$ with $\E V=x$ exactly. It makes one call to $\Round$, $O(ds/r)$ calls to $\Mem$
in expectation, and
\[
   \E_{x}[T]\ =\ O\!\left(\frac{d^{7/2}}{r\,\dmin^{2}}\log\frac{d}{r\dmin}\right)
   \ \subseteq\ O\!\left(\frac{d^{7/2}}{\dmin^{3}}\log\frac{d}{\dmin}\right)
\]
input-coin flips in expectation, where $\dmin=\min_i\{x_i,1-x_i\}$ and $r\ge\dmin$ is the radius
of Lemma~\ref{lem:lpcenter}. Its expected running time is polynomial in $d$, $1/\dmin$, the
encoding length of $(W,b)$ and the cost of $\Round$; for an explicitly given slice, $\Mem$ is the
test $Wz=b$, $0\le z\le1$\revcolor{,} and $\Round$ \revcolor{runs in polynomial time}
\cite[Thm.~6.5.11]{GLS1988}.
\end{theorem}

For perfect matchings the theorem gives the following bound.

\begin{corollary}[Perfect matchings]\label{cor:matching}
Let $n\ge2$. \revcolor{When run} on $P=\Bn$, $\Univ$ is a single algorithm that, for every $X\in\Bnp$, terminates
almost surely and outputs a permutation matrix $M$ with $\E M=X$ exactly, using
\[
   \E_{X}[T]=O\!\left(\frac{n^{8}}{\dmin^{2}}\log\frac{n}{\dmin}\right),
   \qquad \dmin=\min_{ij}X_{ij},
\]
input-coin flips in expectation and expected time polynomial in $n$ and $1/\dmin$.
\end{corollary}

\begin{proof}
The polytope $\Bn=\An\cap[0,1]^{n\times n}$ is an explicitly given affine slice of the cube with
$d=n^{2}$, so $s=n$; no entry is constant on $\Bn$, its vertices are the permutation matrices,
and $\Int(\Bn)=\Bnp$. For $X\in\Bnp$, $1-X_{ij}=\sum_{k\ne j}X_{ik}\ge\min_{kl}X_{kl}$, so
$\min_{ij}\{X_{ij},1-X_{ij}\}$ is the smallest entry of $X$. In the linear program of
Lemma~\ref{lem:lpcenter} every row of $z$ sums to one and so has an entry at most $1/n$, and $U$
attains $t=1/n$; hence $r=1/n$. Substituting $d=n^{2}$ and $r=1/n$ in
Theorem~\ref{thm:universal} gives the flip bound, and the running-time bound applies because
$\Bn$ is explicitly given.
\end{proof}

At the uniform matrix $U$, where $\dmin=1/n$, Corollary~\ref{cor:matching} gives $O(n^{10}\log n)$
flips, while the race of \cite{NPS2023} spends at least $n^{2n-1}$ \revcolor{flips in expectation} there
(Section~\ref{sec:intro}). \revcolor{This bound follows directly from the general theorem, and we have
not tried to optimize the polynomial.}

The rest of the section proves Theorem~\ref{thm:universal}. Section~\ref{sec:pilot} analyzes the
pilot, and Section~\ref{sec:univ-proof} follows the proof of Theorem~\ref{thm:interior}.

\subsection{The pilot}\label{sec:pilot}

The analysis of the pilot needs two binomial tail bounds.

\begin{lemma}[Binomial tails]\label{lem:bintail}
Let $Z$ be a sum of independent Bernoulli \revcolor{random} variables\revcolor{, and let $\mu=\E Z$}. Then
$\Prb[Z\ge h]\le(e\mu/h)^{h}$ for every integer $h\ge1$, and $\Prb[Z\le\mu/2]\le e^{-\mu/8}$.
\end{lemma}

\begin{proof}
Let $Z=\sum_{\revcolor{i=1}}^{\revcolor{N}}\xi_i$ with independent $\xi_i\in\{0,1\}$, $\E\xi_i=p_i$, $\mu=\sum_ip_i$. For the
upper tail, $\Prb[Z\ge h]\le e_h(p_1,\dots,p_N)\le\mu^{h}/h!\le(e\mu/h)^{h}$, where $e_h$ is the \revcolor{$h$-th}
elementary symmetric polynomial and the last step is $h!\ge(h/e)^{h}$. For the lower tail,
\revcolor{Markov's inequality} applied to $e^{-Z}$ gives, with
$\E e^{-Z}=\prod_i\bigl(1-p_i(1-e^{-1})\bigr)\le e^{-\mu(1-e^{-1})}$,
\[
   \Prb[Z\le\mu/2]\le e^{\mu/2}\E e^{-Z}\le\exp\bigl(\mu/2-\mu(1-e^{-1})\bigr)\le e^{-\mu/8},
\]
the last step because $e^{-1}<3/8$ and hence $\tfrac12-(1-e^{-1})\le-\tfrac18$.
\end{proof}

\begin{lemma}[The pilot]\label{lem:pilot}
The pilot terminates almost surely, and with $\bar\lambda:=\E[\lambda]$ and $\varepsilon:=1/64$,
\begin{equation}\label{eq:pilotmain}
   \Prb[\lambda>\dmin]\ \le\ 2\psi\dmin,
   \qquad
   \bar\lambda\ \ge\ \frac{\dmin}{32},
   \qquad\text{hence}\qquad
   \Prb[\lambda>\dmin]\ \le\ 64\psi\bar\lambda .
\end{equation}
Its expected number of input-coin flips is $O\bigl(d\,\dmin^{-1}\log\tfrac1{\psi\dmin}\bigr)$,
and for every rational $\theta\ge2$,
\begin{equation}\label{eq:pilotmoment}
   \E\Bigl[\lambda^{-2}\Bigl\lceil\log_{2}\frac{\theta}{\lambda}\Bigr\rceil\Bigr]
   \ =\ O\Bigl(\dmin^{-2}\log\frac{\theta}{\dmin}\Bigr).
\end{equation}
\end{lemma}

\begin{proof}
\emph{Overshoot.} Fix a stage with $\lambda_{\ell}>\dmin$ and a coordinate $i$ achieving $\dmin$,
say $x_i=\dmin$ (the case $1-x_i=\dmin$ is identical with tails in place of heads). The head
count of coin $i$ has mean $N_{\ell}\dmin=h_{\ell}\dmin/(4\lambda_{\ell})$, so
Lemma~\ref{lem:bintail} gives
\[
   \Prb[\text{stage }\ell\text{ passes}]
   \le\Bigl(\frac{e\,\dmin}{4\lambda_{\ell}}\Bigr)^{h_{\ell}}
   \le\Bigl(\frac{\dmin}{\lambda_{\ell}}\Bigr)^{2}\Bigl(\frac e4\Bigr)^{h_{\ell}}
   \le\psi\frac{\dmin^{2}}{\lambda_{\ell}},
\]
using $\dmin/\lambda_{\ell}\le1$ and $h_{\ell}\ge2$ for the middle step, and
$(e/4)^{h_\ell}\le2^{-h_{\ell}/4}\le\psi\lambda_{\ell}$ for the last, which holds because
$e<3$ and $(3/4)^{4}<\tfrac12$. The event $\{\lambda>\dmin\}$ is contained in the union of these
events, and $\sum_{\ell:\lambda_{\ell}>\dmin}\lambda_{\ell}^{-1}<2/\dmin$ because the reciprocals
double and the largest is below $1/\dmin$. This gives the first bound in \eqref{eq:pilotmain}.

\emph{Typical scale.} Let $\ell_{0}$ be the first index with $\lambda_{\ell_{0}}\le\dmin/8$.
Since $\dmin\le\tfrac12$ we have $\lambda_{0}=\tfrac12>\dmin/8$, so $\ell_{0}\ge1$ and
\begin{equation}\label{eq:ellzero}
   \frac{\dmin}{16}<\lambda_{\ell_{0}}\le\frac{\dmin}{8}.
\end{equation}
At a stage $\ell\ge\ell_{0}$ every one of the $2d$ head and tail counts has mean at least
$N_{\ell}\dmin\ge2h_{\ell}$, so by Lemma~\ref{lem:bintail} and a union bound
\[
   \Prb[\text{stage }\ell\text{ fails}]\le2d\,e^{-h_{\ell}/4}\le2d\psi\lambda_{\ell}
   \le d\psi\le2^{-14}<\varepsilon,
\]
using $e^{-h_{\ell}/4}\le\psi\lambda_{\ell}$, which holds by the definition of $h_{\ell}$
because $\lceil\log_{2}v\rceil\ge\ln v$. Stages are independent, so
\begin{equation}\label{eq:reach}
   \Prb[\text{stage }\ell_{0}+j\text{ is reached}]\le\varepsilon^{j},
\end{equation}
which proves almost sure termination. In particular the pilot stops at some $\ell\le\ell_{0}$
with probability at least $1-\varepsilon$, and then $\lambda\ge\lambda_{\ell_{0}}$; with
\eqref{eq:ellzero},
\[
   \bar\lambda\ \ge\ (1-\varepsilon)\lambda_{\ell_{0}}\ \ge\ \frac{63}{64}\cdot\frac{\dmin}{16}
   \ >\ \frac{\dmin}{32}.
\]
Combining the two displays gives $\Prb[\lambda>\dmin]\le2\psi\dmin\le64\psi\bar\lambda$.

\emph{Cost.} Since $N_{\ell}=(h_{0}+4\ell)2^{\ell-1}$ at least doubles in $\ell$, the total budget
of all stages before $\ell_{0}$ is less than $N_{\ell_{0}}$. By \eqref{eq:reach} the expected
budget from stage $\ell_{0}$ on is at most
\[
   \sum_{j\ge0}\varepsilon^{j}N_{\ell_{0}+j}
   =\frac1{4\lambda_{\ell_{0}}}\sum_{j\ge0}(2\varepsilon)^{j}\bigl(h_{\ell_{0}}+4j\bigr)
   =O\Bigl(\frac{h_{\ell_{0}}}{\lambda_{\ell_{0}}}\Bigr).
\]
Multiplying by $d$ coins and using \eqref{eq:ellzero} and
$h_{\ell_{0}}=O(\log\tfrac1{\psi\dmin})$ gives the stated cost.

\emph{The moment \eqref{eq:pilotmoment}.} Write $L_{\ell}=\lceil\log_{2}(\theta/\lambda_{\ell})\rceil
=L_{0}+\ell$. The quantity $\lambda_{\ell}^{-2}L_{\ell}=4^{\ell+1}L_{\ell}$ increases in $\ell$, so
stops at $\ell\le\ell_{0}$ contribute at most its value at $\ell_{0}$, while by \eqref{eq:reach}
the rest contributes at most
$\lambda_{\ell_{0}}^{-2}\sum_{j\ge0}(4\varepsilon)^{j}(L_{\ell_{0}}+j)=
O(\lambda_{\ell_{0}}^{-2}L_{\ell_{0}})$, the series converging because $4\varepsilon=\tfrac1{16}$.
Now use \eqref{eq:ellzero}.
\end{proof}

\subsection{Proof of Theorem~\ref{thm:universal}}\label{sec:univ-proof}

As in Section~\ref{sec:proofinterior}, the proof collects the facts established along the way, in
three lemmas: one about an adaptive batch, one about the ratio coin, and one about the correction.

\begin{lemma}[The adaptive batch]\label{lem:adaptive}
Let $a=\E[\widehat\alpha]$ and $\bar R=\E[R]$ for one adaptive batch. Then $0<\widehat\alpha\le\tfrac18$,
$a=\bar\lambda/(4s)>0$, and
\begin{align}
 &\tilde Y\in P; \label{eq:ufeas}\\
 &R\in L,\qquad (1-\widehat\alpha)\tilde Y=Y-\widehat\alpha u-R,\qquad
   \E\bigl[(1-\widehat\alpha)\tilde Y\bigr]=x-a\,u-\bar R; \label{eq:uunb}\\
 &|R_i|\le\|R\|_2\le\sqrt d\quad(i\in[d]); \label{eq:ubdd}\\
 &c\,\Prb[R\ne0]\ \le\ \frac{a}{32}. \label{eq:urare}
\end{align}
Its expected number of input-coin flips is
\begin{equation}\label{eq:Cx}
   C_{x}=O\Bigl(\frac{d^{2}}{\dmin^{2}}\log\frac{d}{r\dmin}\Bigr),
\end{equation}
and its expected number of bit operations is polynomial in $d$, $1/\dmin$ and the encoding
lengths.
\end{lemma}

\begin{proof}
Since $\lambda\le\lambda_0=\tfrac12$ and $s\ge1$, $\widehat\alpha\le\tfrac18$, and
$a=\bar\lambda/(4s)$ by the definition of $\widehat\alpha$. Conditionally on $\lambda$ the batch
is the batch of Section~\ref{sec:idea} with $\widehat\alpha$ in place of $\alpha$, on coin flips
independent of the pilot. So the proofs of \eqref{eq:feas}--\eqref{eq:bdd} in
Lemma~\ref{lem:batch} apply verbatim and give \eqref{eq:ufeas}, \eqref{eq:ubdd} and the pointwise
identity in \eqref{eq:uunb}; taking expectations, first conditionally on $\lambda$ and then over
$\lambda$, gives the last part of \eqref{eq:uunb}.

For \eqref{eq:urare} condition on $\lambda$ again. On $\{\lambda\le\dmin\}$,
Lemma~\ref{lem:cuberadius} gives $x\in\Int_\lambda(P)$, so the proof of \eqref{eq:rare} applies
with $\rho=\lambda$, and by \eqref{eq:univ-params2}
\[
   \Prb[R\ne0\mid\lambda]\ \le\ 2d\exp\Bigl(-\frac{m(\lambda)\lambda^{2}}{8s^{2}}\Bigr)
   \ \le\ 2d\cdot\frac{\lambda}{512\,d\,s\,c}
   \ =\ \frac{\lambda}{256\,s\,c}.
\]
On $\{\lambda>\dmin\}$ we assert nothing. Averaging, and using Lemma~\ref{lem:pilot} and
$64\psi=1/(256\,d\,s\,c)$,
\[
   \Prb[R\ne0]\ \le\ \frac{\bar\lambda}{256sc}+\Prb[\lambda>\dmin]
   \ \le\ \frac{\bar\lambda}{256sc}+\frac{\bar\lambda}{256\,d\,s\,c}
   \ \le\ \frac{\bar\lambda}{128\,s\,c}
   \ =\ \frac{a}{32c}.
\]

For the cost, the pilot contributes $O(d\dmin^{-1}\log\tfrac1{\psi\dmin})$ flips by
Lemma~\ref{lem:pilot}, and the batch contributes
\[
   d\,\E[m(\lambda)]
   =O\Bigl(ds^{2}\,\E\Bigl[\lambda^{-2}\Bigl\lceil\log_2\frac{512dsc}{\lambda}\Bigr\rceil\Bigr]\Bigr)
   =O\Bigl(\frac{ds^{2}}{\dmin^{2}}\log\frac{dsc}{\dmin}\Bigr)
\]
by \eqref{eq:pilotmoment}; with $s^{2}\le4d$, $c=ds/r$ and $\log(1/\psi)=O(\log(d/r))$ both are
within \eqref{eq:Cx}. All counts are integers and $\lambda$ is a power of two, so every rational
number formed has encoding length $O(\log d+\log m(\lambda)+\log(1/\lambda)+\bits(u,W,b))$, and
the moments of Lemma~\ref{lem:pilot} bound the expected arithmetic cost.
\end{proof}

The ratio coin of steps (a) and (b) divides by the mean of a coin without a lower bound on it.

\begin{lemma}[Ratio without denominator advice]\label{lem:ratio}
Let $c>0$ be a known rational and let independently reusable coins of means $p\ge0$ and $a>0$ be
available, with the promise $cp/a\le\tfrac18$. Steps (a) and (b) of the correction, run with these
two coins, output a bit of mean exactly $cp/a$, using at most $4K_0(c+1)/a$ calls to the two
coins in total, in expectation. No lower bound on $a$ is used.
\end{lemma}

\begin{proof}
Step (a) returns $1$ with probability
\[
   q=\frac{\tfrac{c}{c+1}p}{\tfrac{c}{c+1}p+\tfrac1{c+1}a}=\frac{cp}{a+cp},
\]
and its number of source calls is geometric with success probability $(cp+a)/(c+1)$, hence at
most $(c+1)/a$ in expectation. By the promise,
$q=\tfrac{cp/a}{1+cp/a}\le\tfrac{1/8}{9/8}=\tfrac19$, so $2q\le\tfrac29<\tfrac14$ and each run of
$\Lin(2,\cdot)$ is within its promise and outputs a bit of mean exactly $2q$. Step (b) therefore
returns $1$ with probability
\[
   \sum_{j\ge1}2^{-j}(2q)^{j}=\sum_{j\ge1}q^{j}=\frac{q}{1-q}=\frac{cp}{a}.
\]
The number of runs of $\Lin(2,\cdot)$ is at most $G$, drawn independently with $\E G=2$; each run
uses at most $2K_0$ flips of the $q$-coin in expectation, and each $q$-coin flip uses at most
$(c+1)/a$ source calls in expectation. Whether a call is made is decided before its own randomness
is drawn, so summing conditional expected costs gives $\E G\cdot2K_0\cdot(c+1)/a\revcolor{{}=4K_0(c+1)/a}$.
\end{proof}

\begin{lemma}[The correction]\label{lem:correctuniv}
Step 2 of $\Univ$ outputs a vertex $C$ of $P$ with
\[
   \E C=u+\frac{\bar R}{a},
\]
and uses at most $4K_0(c+1)/a$ adaptive batches in expectation.
\end{lemma}

\begin{proof}
Fix $i$ and $\sigma$. The coin of the first kind in step (a) is an adaptive batch followed by a
coin of bias $\pos{\sigma R_i}/s$, a rational number in $[0,1]$ by \eqref{eq:ubdd}, so it has mean
$p_i^{\sigma}=\E[\pos{\sigma R_i}/s]$; the coin of the second kind has mean $\E[\widehat\alpha]=a$;
and both are independently reusable because every adaptive batch is fresh. Since
$\pos{\sigma R_i}/s\le\one\{R\ne0\}$ pointwise, \eqref{eq:urare} gives
$c\,p_i^{\sigma}/a\le\tfrac1{32}$, so by Lemma~\ref{lem:ratio} the bit $\xi'$ has mean exactly
$c\,p_i^{\sigma}/a$ and costs at most $4K_0(c+1)/a$ adaptive batches \revcolor{in expectation}. From here the proof of
Lemma~\ref{lem:correct} applies verbatim with $\beta$ replaced by $c/a$: using
$p_i^{+}-p_i^{-}=\bar R_i/s$, $\bar R\in L$ by \eqref{eq:uunb}, and $cr/(ds)=1$, the point handed
to $\Round$ has mean
\[
   u+\frac{cr}{a\,ds}\sum_{i=1}^{d}\bar R_i\,\Pi_L e_i=u+\frac1a\,\Pi_L\bar R=u+\frac{\bar R}{a}.
\]
The anchors are rational points of $\mathbb B(u,r)\cap\aff(P)\subseteq P$, so $\Round$ returns a
vertex, and \revcolor{since $\Round$ is mean preserving, $\E C=u+\bar R/a$}.
\end{proof}

\begin{proof}[Proof of Theorem~\ref{thm:universal}]
\emph{Vertices.} Every point handed to $\Round$ is a rational point of $P$: $\tilde Y$ in step 1
by \eqref{eq:ufeas}, and the anchors in step 2. So the output is a vertex of $P$.

\emph{Exactness.} Conditionally on the adaptive batch of step 1, the algorithm outputs a vertex of
mean $\tilde Y$ with probability $1-\widehat\alpha$ and, with probability $\widehat\alpha$, the
output of a correction that uses fresh randomness and therefore has mean $u+\bar R/a$ by
Lemma~\ref{lem:correctuniv}. Hence, by \eqref{eq:uunb},
\[
   \E V=\E\bigl[(1-\widehat\alpha)\tilde Y\bigr]+\E[\widehat\alpha]\Bigl(u+\frac{\bar R}{a}\Bigr)
       =(x-a\,u-\bar R)+a\,u+\bar R=x .
\]
This uses only the pointwise relation $(1-\widehat\alpha)\tilde Y=Y-\widehat\alpha u-R$ and the
independence of the correction from the first batch, so the correlation between the random reserve
and $\tilde Y$ does no harm.

\emph{Cost and termination.} Step 1 uses one adaptive batch. Step 2 uses at most $4K_0(c+1)/a$
adaptive batches in expectation by Lemma~\ref{lem:correctuniv}, and is entered with probability
exactly $\E[\widehat\alpha]=a$. Each adaptive batch has expected cost $C_x$ regardless of the
history, because whether it is run is decided before its own coins are flipped; so the expected
number of adaptive batches is at most $1+4K_0(c+1)$, the factor $1/a$ cancelling, and
\[
   \E_x[T]\ \le\ \bigl(1+4K_0(c+1)\bigr)\,C_x
   \ =\ O\Bigl(\frac{ds}{r}\cdot\frac{d^{2}}{\dmin^{2}}\log\frac{d}{r\dmin}\Bigr)
   \ =\ O\Bigl(\frac{d^{7/2}}{r\,\dmin^{2}}\log\frac{d}{r\dmin}\Bigr)
\]
by \eqref{eq:Cx} and $s\le2\sqrt d$; the second form of the bound follows from $r\ge\dmin$
(Lemma~\ref{lem:lpcenter}). Each adaptive batch makes one call to $\Mem$, and the algorithm ends
with one call to $\Round$. The expected number of adaptive batches is finite, and the pilot and
the linear factory terminate almost surely, so $\Univ$ terminates almost surely.

\emph{Bit complexity.} Lemma~\ref{lem:adaptive} bounds the expected arithmetic cost of an adaptive
batch and the encoding length of everything it forms. The anchors and the coins flipped in the
correction are rational\revcolor{, with encoding lengths of the same order;} $\Lin(2,\cdot)$ has bounded expected bit cost
(Appendix~\ref{app:scalar})\revcolor{;} and known rational coins are flipped exactly from fair bits at
expected cost polynomial in their encoding length. With the expected number of calls to $\Mem$ and
$\Round$ this gives the running-time bound. For an explicitly given slice the decomposition of
\cite[Thm.~6.5.11]{GLS1988} is polynomial, because separation over $P$ is trivial and the vertices
of $P$ have polynomial encoding length.
\end{proof}

\begin{remark}[Boundary inputs]\label{rem:boundary}
$\Univ$ does not terminate when some $x_{i}\in\{0,1\}$, because the pilot never collects its
required heads or tails there. Strong factories, which terminate everywhere on $P$, exist for
\revcolor{matchings} \cite[\S7]{Multi2023}, and we do not combine that boundary behavior with the interior
flip bound: running a boundary-valid factory in parallel and taking whichever finishes first need
not be correct, since completion time can correlate with the output. If \revcolor{all coordinates of $x$
that lie in $\{0,1\}$ are known}, fixing these coordinates restricts $P$ to a face that is again an affine
slice and contains $x$ in its relative interior, and the theorem applies to that face, every
vertex of which is a vertex of $P$. Knowing only some of them does not suffice, since a single
unknown boundary coordinate stalls the pilot.
\end{remark}

\section*{AI Disclosure}

The proofs in this paper were obtained in a series of conversations with ChatGPT Astra, Claude Opus
and Gemini Argon. The authors then substantially rewrote the proofs with the help of these models,
and take full responsibility for the content of this work. The results were also formalized in Lean 4,
and the formalization is available at \url{https://github.com/radniazadeh/bernoulli-poly.git}.

\appendix
\section{A self-contained linear factory}\label{app:scalar}

\revcolor{For completeness, we give a non-optimized exponent-process implementation of the scalar
primitive of Definition~\ref{def:linear}, which follows the method underlying Huber's factory
\cite{Huber2016}.} Using \cite{Huber2016} instead
improves $K_0$ from $140$ to $38/3$.

\begin{lemma}\label{lem:scalar}
Let $c\ge2$ be rational and suppose we have independent flips of a $p$-coin with the promise
$cp\le\tfrac14$. The procedure $\Amp(c)$ below outputs a bit of mean exactly $cp$. It terminates
almost surely, uses fewer than $140\,c$ input flips in expectation, and has expected additional
computational cost polynomial in $c$ and $\bits(c)$.
\end{lemma}

For $\ell\ge0$ define
\[
   b_{\ell}=1+2^{-\ell-2},\qquad k_{\ell}=16\cdot2^{\ell},\qquad
   c_0=c,\qquad c_{\ell+1}=b_{\ell}c_{\ell}.
\]

\begin{procedure}{$\Amp(c)$}
Initialize $i=1$ and $\ell=0$, and repeat the following \emph{phase}.
\begin{enumerate}[leftmargin=2em,itemsep=1pt,topsep=2pt]
\item While $0<i<k_{\ell}$: flip the $p$-coin. On heads set $i\leftarrow i-1$. On tails draw $G$
  from the geometric distribution on $\{1,2,\dots\}$ with success probability
  $(c_{\ell}-1)/c_{\ell}$ and set $i\leftarrow i-1+G$.
\item If $i=0$, return $1$.
\item Otherwise $i\ge k_{\ell}$. Flip a known coin of mean $b_{\ell}^{-i}$. On tails return $0$;
  on heads set $\ell\leftarrow\ell+1$ and begin the next phase, retaining $i$.
\end{enumerate}
A phase beginning with $i\ge k_{\ell}$ skips its inner loop.
\end{procedure}

\begin{proof}[Proof of Lemma~\ref{lem:scalar}]
\emph{A bounded martingale.} For every $\ell$,
$c_{\ell}\le c\exp\bigl(\sum_{j\ge0}2^{-j-2}\bigr)=e^{1/2}c<2c$, so $q_{\ell}:=c_{\ell}p<2cp\le
\tfrac12$. Attach to the state $(i,\ell)$ the value $q_{\ell}^{\,i}\in[0,1]$. Within a phase write
$q=q_{\ell}$. For a geometric $G$ with success probability $(c_{\ell}-1)/c_{\ell}$ and $q<1$,
\[
   \E[q^{G}]=\frac{\frac{c_{\ell}-1}{c_{\ell}}q}{1-\frac{q}{c_{\ell}}}
            =\frac{(c_{\ell}-1)q}{c_{\ell}-q}=\frac{(c_{\ell}-1)p}{1-p},
\]
using $q=c_{\ell}p$; hence one inner step preserves the value,
\[
   \E\bigl[q^{\,i_{\mathrm{new}}}\mid i\bigr]
   =q^{i-1}\Bigl(p+(1-p)\tfrac{(c_{\ell}-1)p}{1-p}\Bigr)
   =q^{i-1}c_{\ell}p=q^{i}.
\]
At a phase transition the process survives with probability $b_{\ell}^{-i}$ and then has value
$(b_{\ell}c_{\ell}p)^{i}=b_{\ell}^{\,i}q^{i}$, while a killed process has value $0$; the
expectation is again $q^{i}$. The absorbing state $i=0$ has value $1$ and returns $1$; death has
value $0$ and returns $0$; the initial value is $(c_0p)^{1}=cp$. So these values form a bounded
martingale started at $cp$, and once the procedure is known to terminate almost surely, optional
stopping identifies the output mean as $cp$.

\emph{Expected number of flips.} Inside a phase the expected increment of $i$ is
\[
   d_{\ell}=-1+(1-p)\frac{c_{\ell}}{c_{\ell}-1}=\frac{1-c_{\ell}p}{c_{\ell}-1}
   \ \ge\ \frac{1/2}{2c}=\frac1{4c},
\]
using $c_{\ell}p\le\tfrac12$ and $c_{\ell}-1<2c$. Run the inner chain from $i_0<k_{\ell}$ until it
hits $0$ or reaches $k_{\ell}$, at time $\sigma$. At an upper exit, writing $i_{\sigma}=i-1+G$
conditioned on $G\ge k_{\ell}-i+1$, the memoryless property gives
$\E[i_{\sigma}-k_{\ell}]\le\frac{c_{\ell}}{c_{\ell}-1}-1\le1$ because $c_{\ell}\ge c\ge2$; hence
$\E[i_{\sigma}]\le k_{\ell}+2$ in all cases. Applying the drift identity
$\E[i_{\sigma\wedge n}]-i_0=d_{\ell}\E[\sigma\wedge n]$ at a deterministic truncation $n$ --- for
which the terminal value is still at most $k_{\ell}+2$ in expectation, since non-exited
trajectories sit below $k_{\ell}$ --- and letting $n\to\infty$ by monotone convergence,
\begin{equation}\label{eq:phasecost}
   \E[\text{inner steps in phase }\ell]\ \le\ \frac{\E[i_{\sigma}]}{d_{\ell}}
   \ \le\ 4c\,(k_{\ell}+2).
\end{equation}
At a phase transition $i\ge k_{\ell}=16\cdot2^{\ell}$, and $\log(1+v)\ge v/2$ for $v\in[0,1]$
gives $\log b_{\ell}\ge2^{-\ell-3}$, so $b_{\ell}^{-i}\le e^{-2}<\tfrac14$. Conditionally on any
history the next phase is reached with probability at most $\tfrac14$, so phase $\ell$ is reached
with probability at most $4^{-\ell}$. With \eqref{eq:phasecost},
\[
   \E[\#\ p\text{-coin flips}]\ \le\ \sum_{\ell\ge0}4^{-\ell}\,4c\,(16\cdot2^{\ell}+2)
   \ =\ 4c\Bigl(32+\tfrac83\Bigr)=\frac{416}{3}c\ <\ 140\,c .
\]
In particular the expected number of flips is finite and the procedure terminates almost surely,
completing the martingale argument. When $p=0$ no downward step is possible, $i$ never reaches
$0$, and the output is always $0=cp$; the drift is still at least $1/(4c)$, so the same cost and
termination bounds apply.

\emph{Bit cost.} All simulated probabilities are rational. A geometric draw needs at most two
known-coin flips in expectation, its success probability being at least $\tfrac12$. The terminal
$b_{\ell}^{-i}$-coin is realized by flipping $b_{\ell}^{-1}$-coins until the first tail or until
$i$ heads occur, at expected cost $\sum_{j\ge0}b_{\ell}^{-j}=b_{\ell}/(b_{\ell}-1)=2^{\ell+2}+1$;
these costs remain summable against $4^{-\ell}$. The \revcolor{rational number} $c_{\ell}$ has
$O(\bits(c)+\ell^{2})$ bits, counters before an update satisfy $i<k_{\ell}$, and geometric
overshoots have uniformly exponential tails. Standard integer arithmetic therefore gives expected
polynomial bit cost.
\end{proof}

\end{document}